\documentclass[prl,aps,superscriptaddress,twocolumn]{revtex4-2}

\usepackage{amsmath}
\usepackage{amssymb}
\usepackage{graphicx}
\usepackage{mathrsfs}
\usepackage{siunitx}
\usepackage[version=4]{mhchem}
\begin{document}

\title{Multi-particle gates on driven one-dimensional paths: probing deep traps}

\author{Harsh Jain}
\email{harshjainldh@gmail.com}

\affiliation{Department of Condensed Matter Physics and Materials Science, Tata Institute of Fundamental Research, Mumbai 400005, India}
\affiliation{National Center for Biological Sciences, Tata Institute of Fundamental Research, Bengaluru}

\author{Shankar Ghosh}
\affiliation{Department of Condensed Matter Physics and Materials Science, Tata Institute of Fundamental Research, Mumbai 400005, India}

\author{Archishman Raju}
\affiliation{National Center for Biological Sciences, Tata Institute of Fundamental Research, Bengaluru}


\begin{abstract}

We study single-file transport of driven overdamped colloidal particles on a periodic path with deep potential wells. In the small trap limit (i.e., trap size smaller than particle size), the  particle current transitions from zero to finite as the number of particles on the path exceeds a critical number $n_c$. Beyond this threshold, $n_c$ particles cluster behind the trap, demonstrating collective correlated motion. The remaining `extra' particles circulate, giving a finite current. We study this phenomenon numerically using overdamped Brownian dynamics simulations, and present an experimental realization of this behaviour for micron-scale colloidal particles driven in an optical vortex. Using our experimental observations, we present results characterizing  potential wells as deep as several hundred $k_BT$.  
\end{abstract}


\maketitle

Particles forbidden from overtaking via hard-core interactions on a one-dimensional path have been studied extensively in the context of
the asymmetric simple exclusion process (ASEP)~\cite{Derrida1998, Kumar2020TASEP}.
Driven single-file diffusion of interacting colloidal particles has
been studied theoretically~\cite{jain2007driving, chaudhuri2015pumping}, with
experimental realizations exploring transport across structured optical
landscapes~\cite{Bechinger_Seifert_2007characterizing, juniper2016colloidal,
	kumar2013transition, haider2016enhanced}. 
One-dimensionality is essential for control: in higher dimensions,
multiple escape paths with varying potential landscapes become accessible to the system, complicating both the dynamics and its theoretical description~\cite{satija2020broad}. Such driven 1D systems have been theoretically modelled as Brownian particles in tilted
periodic potentials~\cite{Stratonovich1967, risken1989fokker}.
	In this paper, we present
particle-gate-like behavior that emerges in driven single-file
diffusion in the presence of deep traps. In the presence of multiple interacting particles, a critical cluster size $n_c$ is required for finite current
— a collective, non-equilibrium effect that simultaneously provides a
means to characterize deep traps. Measuring size or depth of such traps using Kramers' escape-rate theory or umbrella sampling \cite{Kramers1940, kastner2011umbrella} would otherwise remain inaccessible at experimental timescales.

\begin{figure}[ htbp]
    \centering
    \includegraphics[width=1.04\linewidth]{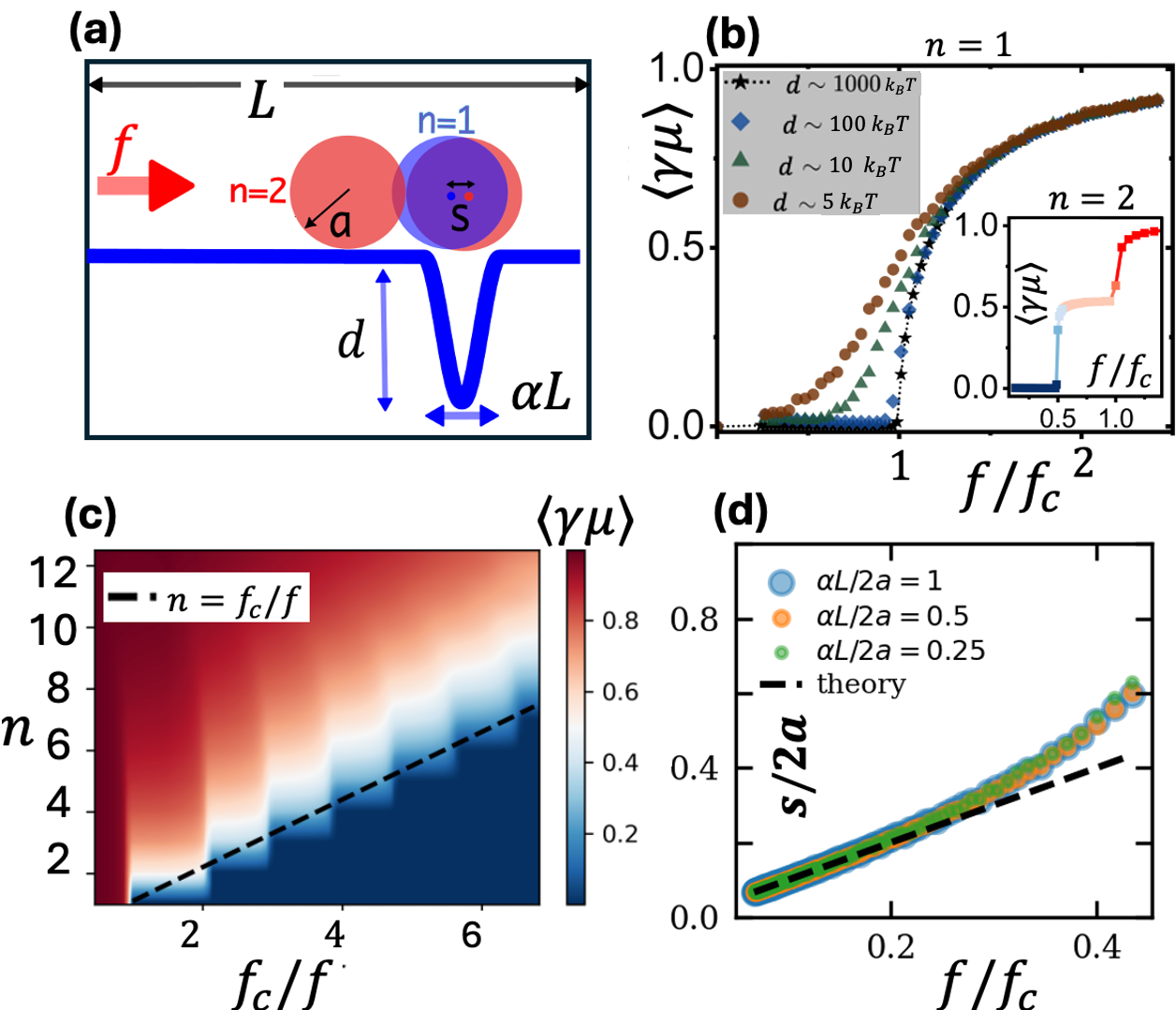}
    \caption{(a) shows a periodic potential well (blue) with potential depth ${d}$ and trap size $\alpha L$, periodicity $L$ and a driving force $f$ along the path. The blue particle represents $n=1$  and the two red particles represent $n=2$ on the path. (b) shows the variation of particle's transport efficiency  $\langle \gamma \mu \rangle$ versus $f/f_c$ for a single particle and various potential well depths $\mathrm{d}$. Inset shows the transport efficiency for $n=2$ and $d=1000 k_BT$. (c) shows the phase diagram for current $\langle \gamma \mu\rangle$ with $n$ particles on the vertical axis and $f/f_c$ on the horizontal axis for $\mathrm{d}= 1000~k_BT$. Dashed line is $n=f_c/f$. (d) shows the normalized shift $s/2a$ in the position of the particle at the edge as a function of $f/f_c$.}
    \label{fig:KramersNonEqm}
\end{figure}

Consider a potential well on a periodic domain of length $L$ characterized by a size parameter $\alpha$ such that the size of the trap is $\alpha L$ and the depth of the trap is $d$ in units of $k_B T$ where $k_B$ is the Boltzmann constant and $T$ is the temperature.  Fig.~\ref{fig:KramersNonEqm}(a) shows such a potential well with a spherical particle of radius $a$ trapped as shown in blue.
We model the potential well using a cosine function. If $x=0$ is assumed to be the starting point of the trap, $U(x) = (d/2)\cos\left(2\pi x/\alpha L\right)$ for  $0 < x < \alpha L$ inside the trap, and the potential is assumed to be constant outside the trap i.e. $U(x) = d/2$. 
In the presence of a weak external force $f$, the non-equilibrium potential is $ \widetilde{U}(x) = d/2\cos\left(2\pi x/\alpha L\right) - f x$;  $x \in [0, L)$. 
By finding the extrema of the potential $\widetilde{U}(x)$, we can conclude that a single particle will remain trapped for $f < f_c$, in the absence of noise where $f_c=\pi\,d /\alpha L$[see Supp. Fig. 1].

Fig.~\ref{fig:KramersNonEqm}(b) shows the numerical solution to the overdamped Langevin equation for the dynamics of a \SI{2}{\micro\meter} diameter colloidal particle in the non-equilibrium periodic potential where $ d=5 k_B T, 10 k_B T, 100 k_B T$ and $1000 k_B T$.  The overdamped equation of motion can be written as 
$
	\gamma \dot{x} = -{\partial \widetilde{U}}/{\partial x} + \eta(t)$, 
where $\gamma$ is the drag coefficient, and $\eta(t)$ represents Gaussian white noise with zero mean and correlation $\langle \eta(t)\eta(t') \rangle = 2\gamma k_B T \delta(t-t')$. Here, the normalized mobility $\langle \mu \rangle = \langle v \rangle/f$  where  $\langle v \rangle$ is the mean drift velocity of the particle on the path. $\gamma \mu$ characterizes the transport efficiency of the system  and saturates to 1 for $f \gg f_c$.  
This problem for a single particle has a known analytical solution (Ref.~\cite{risken1989fokker}) which matches our numerical result. We find that the transition to finite transport efficiency becomes sharper as we have a deeper trap, and therefore the transition becomes more deterministic and less stochastic. Note that the transport efficiency is zero for a single particle in a deep trap for $f<f_c$. We define $n$ as the number of particles.
However, if we have more than one particle in this system interacting via hard-core repulsions, the particles can start accumulating behind the trap as shown in  Fig.~\ref{fig:KramersNonEqm}(a) for $n=2$ with two red particles. The inset in Fig.~\ref{fig:KramersNonEqm}(b) shows that the transport efficiency is greater than zero for $n=2$ even for a deep trap with $d=1000 k_BT$ and $f<f_c$.
 
Fig.~\ref{fig:KramersNonEqm}(c) shows the numerical solutions for  $\gamma \mu$ in this system with a density plot for a deep trap with  $d=1000 k_BT$ and $n\geq1$. We plot $n$ versus $f_c/f$ and find the saturation of $\gamma \mu\approx 1$  even when $f/f_c<1$  (or $f_c/f > 1$ as shown in the plot) for higher values of $n$ along the vertical axis. This transition of $\gamma \mu$ to non-zero values happens at a critical number of particles $n_c \approx \lfloor f_c/f \rfloor$ as shown via the black dashed line. 
This approximation for $n_c$ is valid in the limiting scenario of small potential traps, i.e. $\alpha L \leq 2a$.  This result emerges from collective behavior among the particles that form a one-dimensional cluster at the trap location. As the driving forces exerted on multiple particles add together, a finite particle current starts appearing in the system above the critical particle number $n_c$ for a given $f_c/f$. We find this transition from blue to red along the y-axis to be sharp and therefore compare this with a particle gate. In the small trap limit, the motion among the particles in the cluster is also highly correlated, as we discuss later. The role of noise is more significant in the high $n$ and small $f$ regime as demonstrated by the larger width of the white phase boundary region.
Another interesting observation from Fig.~\ref{fig:KramersNonEqm}(a) is that as the number of particles in the system changes from $n=1$ (blue) to $n=2$ (red), the ``leading" particle in the potential well, at the front of the trapped cluster, shifts by a small length $s$.  Fig.~\ref{fig:KramersNonEqm}(d) shows the plot for this normalised shift $s/2 a$ as this transition from $n=1$ to $n=2$ occurs. 
In the low-force regime ($f \ll f_c$), we derive the position shift of the leading particle upon the addition of a second particle. For a single particle in the tilted cosine potential, force balance yields $f = k_{\text{trap}}x_1$, where $k_{\text{trap}} = 2\pi^2 d/(\alpha^2 L^2)$ is the harmonic trap stiffness near the potential minimum. When a second particle is introduced to the left, the total driving force becomes $2f$, balanced at a new equilibrium position: $2f = k_{\text{trap}}x'$. The position shift of the top particle is thus $s = x' - x_1 = f/k_{\text{trap}}$, yielding the linear scaling relationship
 \begin{equation}
 \frac{s}{2a} = \frac{\alpha L}{4 \pi a} \frac{f}{f_c}
 \label{eqn:shiftForce}
 \end{equation}
 This prediction agrees quantitatively with our numerical simulations in the small $f$ regime where $f/f_c < 0.3$, with deviations at higher forces arising from nonlinear potential corrections as we transition from zero to finite current. The shift $s$ remains the same with the addition of every extra particle when $f\ll f_c$ . 
 
In the next section, we will present how these observations enable us to measure deep potential wells in experimental scenarios. The experimental system we present is for a driven colloidal particle in an optical vortex near a charged glass coverslip. 







\section{Experimental Setup}

\begin{figure}[ htbp]
    \centering
    \includegraphics[width=.99\linewidth]{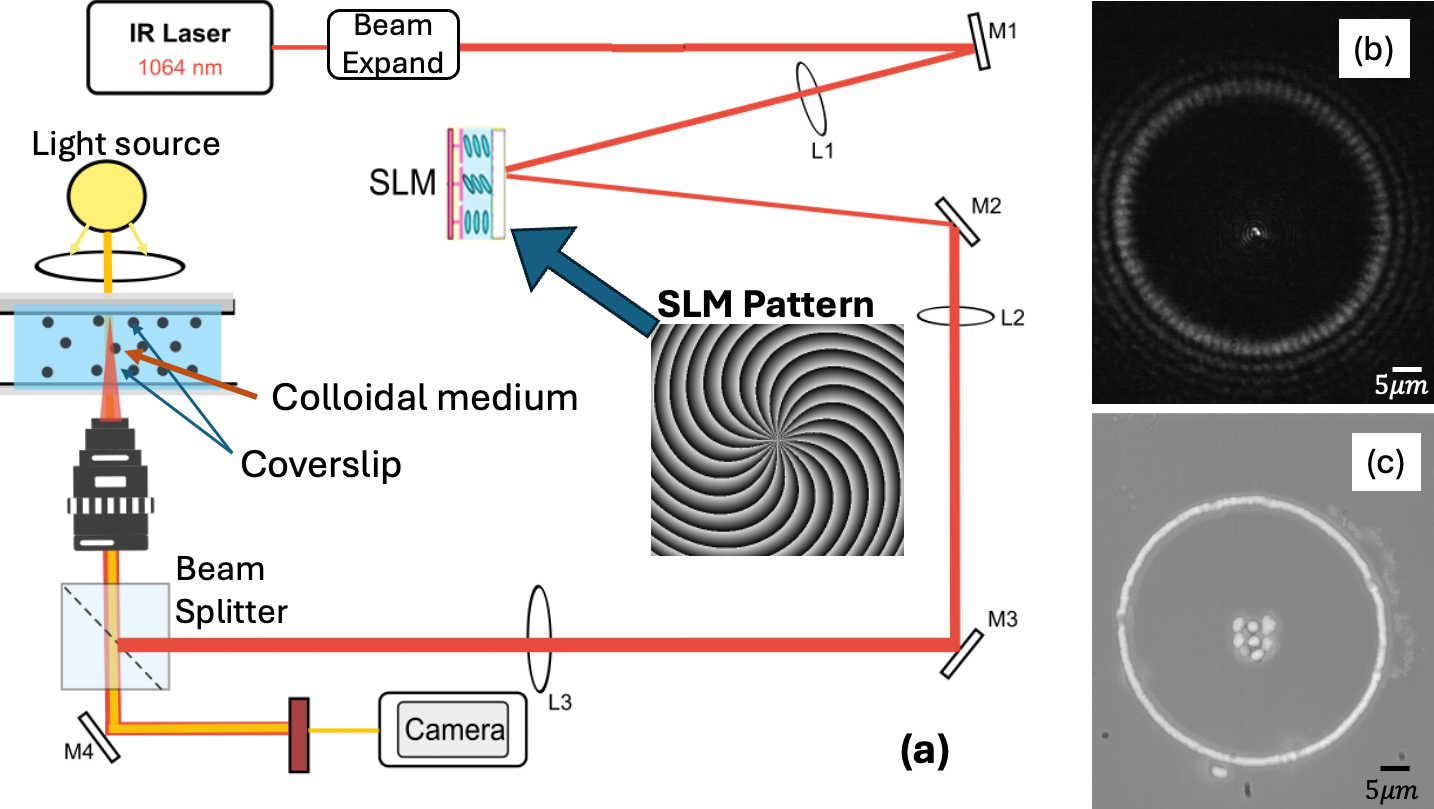}
    \caption{(a) shows the experimental setup for the optical tweezer consisting of a spatial light modulator (SLM), mirrors M1-3 and lenses L1-3, along with the phase pattern of Laguerre Gaussian beam  projected on the SLM. (b) shows the optical vortex seen by the camera reflected by the coverslip. (c) shows a long exposure image (exposure time $\sim 10s$) for particles moving along the circular trap generated by the optical vortex }
    \label{fig:ExptSetup}
\end{figure}


Our experimental setup for trapping colloidal particles along a periodic path is shown in Fig.~\ref{fig:ExptSetup}(a). We generate an optical vortex using single mode output (Gaussian $TEM_{00}$)
from an Nd:YAG laser (Coherent Vector 1064-4000M) with $\lambda=\SI{1064}{\nano\meter}$ and an output power $\sim 1W $ which is current-controlled. 
The laser illuminates the back focal plane of a 60x oil-immersion microscope objective with a Numerical Aperture of 1.3 after passing through a spatial light modulator (SLM, Hamamatsu LCOS 800x600). The optical path is shown in Fig.~\ref{fig:ExptSetup}(a). Through wavefront modulation via the SLM with a phase pattern $e^{i\ell\phi}$, where $\phi$ is the azimuthal angle in the plane of the beam and $\ell$ is the topological charge, we can generate a circular optical trap also known as an optical vortex as shown in Fig.~\ref{fig:ExptSetup}(b)\cite{COULLET1989403, Heckenberg:92,PhysRevA.45.8185}.  
Hotspots, i.e., aberrations along the path of optical vortices \cite{grier2003structure} have been reported previously in the literature and were resolved by the use of axicons\cite{Dholakia2013PerfectVortex}. 
We coupled the phase mask of an axicon directly onto the SLM along with a phase mask for a Laguerre-Gaussian beam with topological charge $\ell=40$. The generated phase pattern looks similar to the SLM pattern shown in Fig.~\ref{fig:ExptSetup}(a). This generates a uniform intensity of light along the circular path and we obtain optical vortices without any optical ``hot spots". As a result, we observe colloidal particles moving around in the trap without getting trapped at a specific location as shown in Supplementary Video 1 for particles of radius $\SI{2}{\micro\meter}$. The algorithm for generating this pattern can be found at \cite{HarshGithubSLM} and calculation of the phase pattern is discussed in Supplementary section II. 
The applied SLM pattern's Fourier transform was obtained as a circular optical vortex in the focal plane shown in Fig.~\ref{fig:ExptSetup}(b). The image was captured by a CCD camera with exposure time of 10 milliseconds.
We trap spherical polystyrene colloidal particles (Bangs Laboratories, latex microspheres) with radius $a \sim 0.3-\SI{2}{\micro\meter}$ in our experiments. The polystyrene particles are suspended inside de-ionised water and constrained between two glass coverslips in our setup. 
Planar gradient forces confine colloidal particles along the circular path of radius $R\approx \SI{22}{\micro\meter}$, while radiation pressure from the laser pushes particles vertically towards the top coverslip. 
Orbital Angular momentum is transmitted onto the colloidal particles as each scattered photon imparts a momentum $\ell \hbar$ to the particle where $\ell$ is encoded in the SLM's phase pattern \cite{Dholakia2013PerfectVortex,grier2003structure,PhysRevA.45.8185}.
Effectively, we obtain a non-equilibrium driving force $f$ along the tangential direction. Long exposure images of the particles with diameter \SI{2}{\micro\meter} driven along the circular trapped path are shown in Fig.~\ref{fig:ExptSetup}(c).

Here, $\ell = 40$ and therefore, each photon imparts tangential momentum $\ell\hbar/R$ at ring radius R = \SI{22}{\micro\meter} and the circumferential path width $w \approx \SI{2}{\micro\meter}$. With the laser power set at 1~W, 0.3 W of power was found to reach the objective as measured with a laser power meter. Here, the dominant loss arises from the liquid crystal SLM with  attenuation $\sim 50\%$ and the remainder attenuated by the steering mirrors and lenses. Approximately 0.2~W is concentrated in the primary vortex ring as estimated from intensity profile of the trap. The vortex ring occupies an area $2\pi R w$ and so the trapped particle experiences an intensity $I \sim 10^9$~W/m$^2$. For a polystyrene sphere (refractive index, $n_p = 1.59$) in water($n_w = 1.33$) with radius $a \approx \SI{1} {\micro\meter}$,  the particle geometrically intercepts $\sim  10^{16}$ photons per second through its cross-section which would yield a total radiation pressure force estimate on the particle $f_{total} \sim 10$~pN assuming complete momentum transfer.  This radiation pressure force can be decomposed into a tangential (azimuthal) component that drives orbital motion, and the remaining axial (vertical) component presses the particle towards the top coverslip \cite{Padgett1995,Leach2006,BohrenHuffman1983}. The helical phase factor $e^{i\ell\phi}$ advances by $2\pi\ell$ over one 
full azimuthal circuit of circumference $2\pi R$, giving a local azimuthal 
phase gradient of $\ell/R$. Since the total phase advance per 
metre along the propagation direction is $2\pi n_w/\lambda$, the sine of the 
cone half-angle $\theta$ is simply the ratio of these two rates
$	\sin\theta = \ell\lambda/2\pi n_w R	\approx 0.23$. Therefore, we estimate the tangential force $f = f_{total} \sin\theta\sim \SI{2}{\pico\newton}$ as an upper bound.  These calculations are detailed in Supplementary 
Section~III.

We used deionised water as the suspending medium for the polystyrene colloidal 
particles in our experiments. In aqueous solution at neutral pH, the 
polystyrene microspheres acquire a negative surface charge from the ionisation 
of sulfate groups
\cite{bangs_tn100}, while the glass coverslip 
develops a negative surface charge from the deprotonation of surface silanol 
groups
\cite{behrens2001charge}. 
The resulting 
repulsive electrical double-layer interaction 
constitutes the Derjaguin--Landau--Verwey--Overbeek (DLVO) 
potential between the particle and the wall~\cite{israelachvili2011intermolecular}. 
At the low ionic strengths in deionised water, the Debye 
screening length is large ($\gtrsim \SI{100}{\nano\meter}$), resulting in 
a long-ranged electrostatic repulsion and a  DLVO barrier 
that prevents adhesion of the particle to 
the coverslip. The equilibrium 
hovering height of the particle is set by the balance between this repulsive 
DLVO interaction and the vertical scattering force exerted by the laser 
beam, which pushes the particle towards the coverslip~\cite{jones2015optical, harada1996radiation}.
Real glass surfaces are not, however, electrostatically uniform. Spatial 
fluctuations in the surface charge density arise from silanol clustering~\cite{behrens2001charge, sonnefeld1995}. 
Surface morphology and charge based studies have established that such 
heterogeneities exist on length scales of around several 
hundred nanometers~\cite{duffadar2009impact,kozlova2007,kalasin2015,siretanu2014,walz1998}. At these localised charge 
``pockets'', the DLVO barrier is reduced or eliminated, allowing the colloidal particles 
to encounter them laterally as they move along our non-equilibrium periodic path parallel to the coverslip, with depth of several hundred to 
a few thousand $k_B T$ ~\cite{ruckenstein1976, israelachvili2011intermolecular}. The 
micron-scale particle does not resolve individual nanoscale patches but 
rather integrates the surface charge over a lateral ``zone of influence'' 
of size comparable to the particle size, so that the 
effective lateral extent of the resulting potential well is set by the 
convolution of the patch distribution with the length 
scale $a$ \cite{duffadar2009impact,bendersky2011,bhattacharjee1997}.
In our experiments, we observed that increasing the ionic strength of the water medium by adding $\ce{NaCl}$ salt leads to rapid 
adhesion of the particles to the coverslip, consistent 
with classical DLVO predictions related to lowering of the Debye length and the DLVO barrier~\cite{israelachvili2011intermolecular,elimelech1995}.  
We also observed that for a small-sized particle with diameter $\approx \SI{0.6}{\micro\meter}$, the zone of influence is small and because of the roughness of the potential energy landscape, the particle experiences a larger number of traps (shown in Supp. Video 2). As a result, the mobility is significantly reduced. On the other hand, larger particles lead to higher mobility and much more averaged behavior leading to interesting observations as we report in the next section.

\section{Experimental Observations}

\begin{figure}[ htbp]
	\centering
	\includegraphics[width=.99\linewidth]{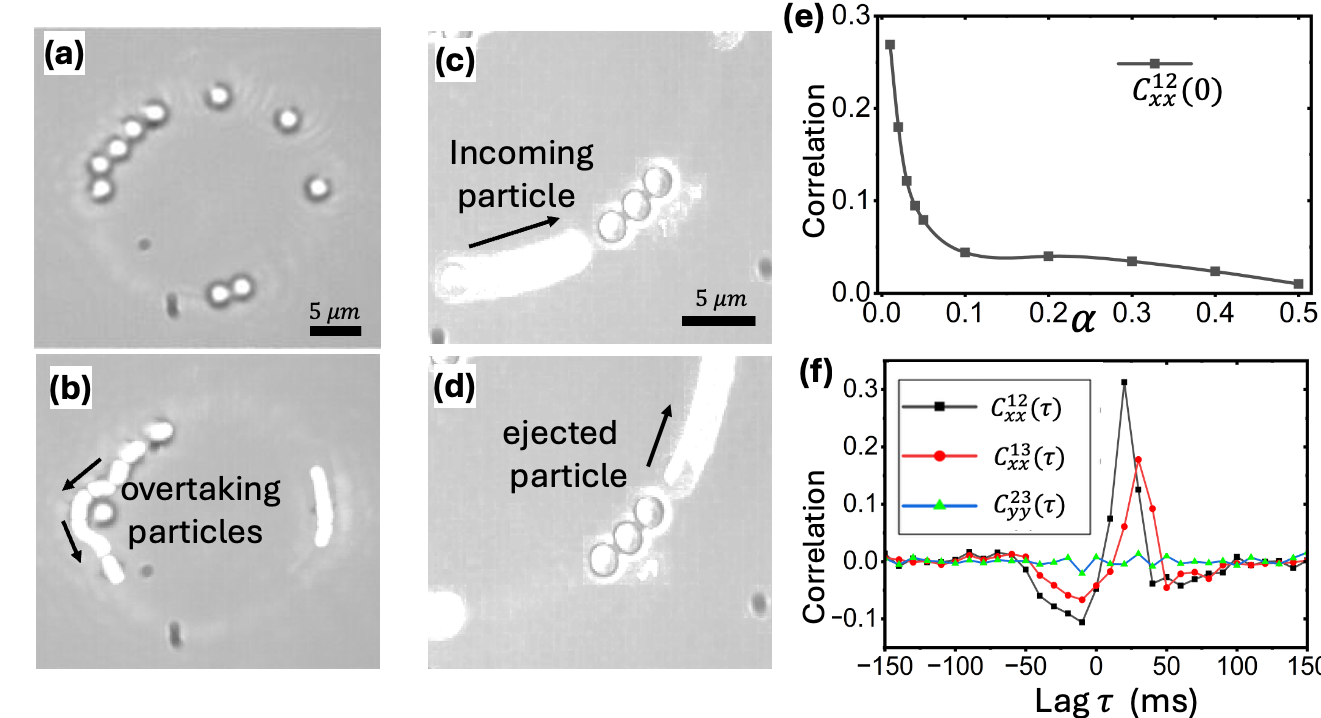}
	\caption{(a) shows a snapshot of \SI{2}{\micro \meter} particles on a circular trap. (b) shows a long exposure shot of particles overtaking a trapped particle in a wide trap. (c) shows long exposure image of a \SI{2}{\micro \meter} particle moving towards a cluster of 3 trapped particles, and (d) shows emission of a particle from the `particle gate'  after collision (e) shows the simulation results for $C_{xx}^{12}(0)$ versus $\alpha$ for particle positions inside the cluster (f) shows the correlation at  lag $\tau$ for experimental data.}
	\label{fig:Images}
\end{figure}

Fig.~\ref{fig:Images} shows colloidal particles moving on the optical vortex path and getting stuck in deep potential wells that result from charge heterogeneities.
Fig.~\ref{fig:Images}(a) shows the snapshot of particles of diameter \SI{2}{\micro\meter} on a wider optical vortex circumferential path with $R\approx \SI{10}{\micro \meter}$,  with exposure time of $\sim \SI{10}{\milli\second}$  and (b) shows a longer exposure time of $\sim \SI{1}{\second}$ in the same scenario. The particles not stuck in a trap are able to overtake a particle stuck in the trap on the circumferential path with the path width $w\gtrsim 2a$. Fig.~\ref{fig:Images}(c-d) shows  particles with diameter $\approx \SI{2}{\micro \meter}$ on a narrower section of the circumferential path with $R\approx \SI{22}{\micro\meter}$ where $w\lesssim 2a$. The particles are unable to overtake each other along the path and therefore get trapped. The leading particle is stuck in the trap and shifts as more particles cluster behind it similar to our model in Fig.~\ref{fig:KramersNonEqm}. In Fig.~\ref{fig:Images}(c), the trap can be characterized with $n_c = 3$. This can be inferred since, when a fourth incoming particle joins the cluster, one particle at the front instantly leaves the cluster as shown in Fig.~\ref{fig:Images}(d).

Another interesting observation we make here is that the motion of the $n$ particles in the cluster is highly correlated along their direction of transport. We extend our simulations to include particle interactions via the Weeks-Chandler-Anderson (WCA) interaction potential \cite{Allen1987}. Details of the simulation can be found in Supplementary Section VI. We plot the correlation coefficient $C_{xx}^{12}$ in our simulations as a function of the size parameter $\alpha$, where $x$ is the direction of transport and $1, 2$ index the first and second particle in the cluster from the front. Evidently, small traps produce correlated movement between the two particles. The normalised cross-correlation of the displacements $\delta x_i =x_i(t + \delta t) - x_i(t)$ of particles at time step $\delta t$  for particle index  $i$ and $j$ at lag $\tau$ is defined as
\begin{equation}
	C^{ij}_{xx}(\tau) = 
	\frac{\langle \delta x_i(t)\,\delta x_j(t+\tau)\rangle}
	{\sqrt{\langle\delta x_i^2\rangle\langle\delta x_j^2\rangle}},
	\label{eq:corr}
\end{equation}
By construction, $C^{ii}_{x x}(0) = 1$. 

For traps of small sizes, represented by $\alpha < 0.1$, the correlation coefficient  $C_{xx}^{12}$ increases sharply demonstrating collective motion among the particles in the cluster. For large traps, no such correlation is observed and correspondingly, the small trap limit result $n_c = \lfloor f_c/f\rfloor$ no longer holds.  

This numerical result is in agreement with our experimental observations for correlation between particle positions in the x-direction as demonstrated in Fig.~\ref{fig:Images}(f) where $C_{xx}^{12}(\tau), C_{xx}^{13}(\tau)$, and $C_{yy}^{23}(\tau)$ are plotted against the lag time $\tau$. 



\begin{figure}[ htbp]
	\centering
	\includegraphics[width=.99\linewidth]{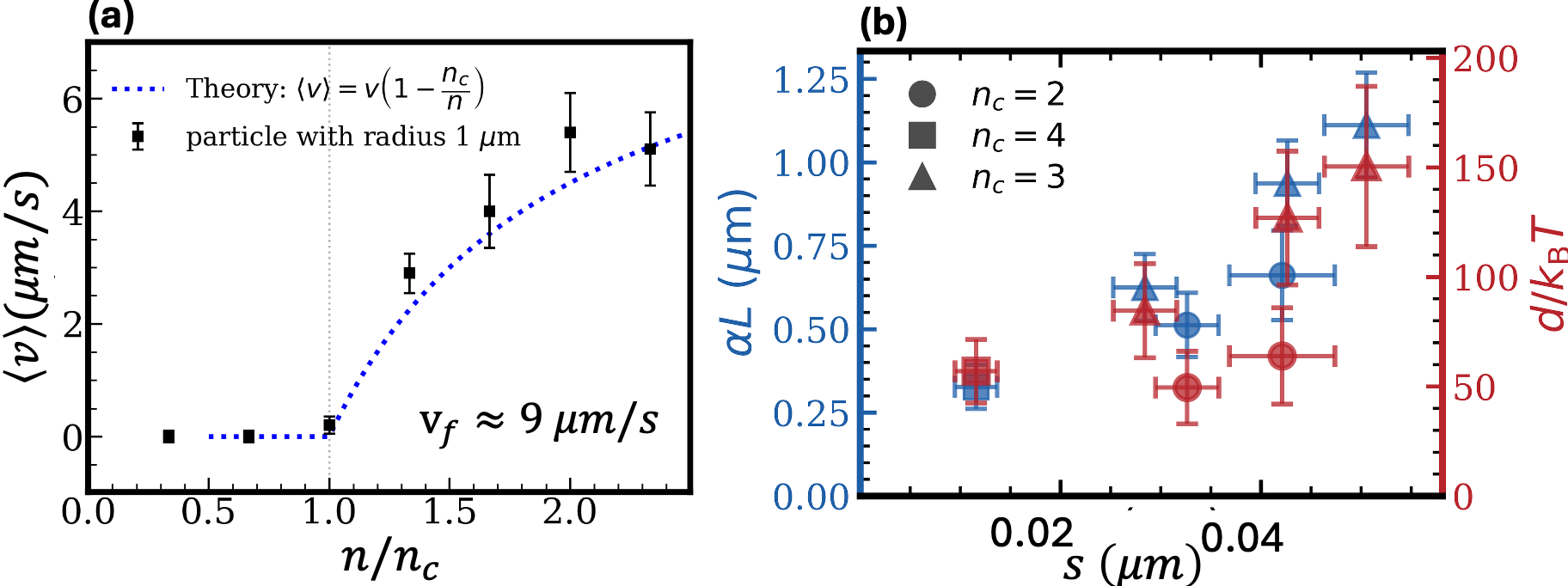}
	\caption{ (a) shows average velocity $\langle v \rangle$ for particles of diameter \SI{2}{\micro\meter} versus $n/n_c$. (b) shows the measured shift $s$ versus trap size $\alpha L $ (left,blue) and normalized depth $d/k_BT$ for the same value of driving force $f$, for $n_c = 2,3,4$ and at different trap locations. 
	}
	\label{fig:particle-gate-measure}
\end{figure}

The average velocity $\langle v \rangle$ of the colloidal particle trapped inside the cluster of $n_c$ particles is almost zero at experimental timescales. As the number of particles $n$ is increased so that $n>n_c$, a critical number of $n_c$ particles continue to be stuck in the cluster with $\langle v \rangle = 0$ and $n-n_c$ extra particles freely move outside the trap with the velocity of the free particle $v_f$. 
We can therefore estimate the average velocity for all the $n$ particles as $\langle v\rangle = v_f (1-n_c/n)$. The agreement between theory and experimental observations is shown in Fig.~\ref{fig:particle-gate-measure}(a).
The mean speed achieved by a freely orbiting colloidal particle 
along the vortex path is observed to be $\langle v_f \rangle \approx 
\SI{9}{\micro\meter\per\second}$. For a particle confined very close to the coverslip surface, 
the drag coefficient is enhanced beyond the bulk Stokes value by the 
hydrodynamic interaction with the wall. In the near-contact limit 
$h \ll a$, the drag on a sphere translating parallel to a plane wall 
is given as $\gamma = 18\pi\eta a$, 
with an enhancement factor of $ 3$ over the bulk Stokes value of $6\pi \eta a$ \cite{happel1965low,Goldman1967}. The tangential driving force is then
$f = 18\pi\eta a\,\langle v \rangle 
\approx 0.5~\text{pN}.$
This suggests a radiation pressure efficiency of $\sim 25\%$ when compared to the force calculated previously for complete momentum transfer. 
A similar value is obtained from a full Mie-theory calculation for polystyrene colloidal particles at this size and wavelength \cite{BohrenHuffman1983} . The calculations for the Mie scattering and drag enhancement factor are detailed in Supplementary Sections IV and V.


We explore multiple traps at different locations on the glass surface using these control parameters. For each trap, we measure the critical number of particles $n_c$ and the shift $s$ of the leading particle. 
Using our results from Fig.~\ref{fig:KramersNonEqm}, we estimate $f_c$ via the bounds  $n_c f < f_c < (n_c+1)f$. Substituting this into Eq.~\ref{eqn:shiftForce}, we get an estimate for $\alpha L$ and the relation $f_c = \pi d/\alpha L$ allows us to estimate the trap depth $d$.
As an example, for $n_c = 3$ and $f \approx 0.5 pN$, we can estimate $f_c = 1.75 pN \pm 0.25 pN$.  The corresponding shift $s$ of the leading particle with the addition of every extra particle was measured  using image processing in MATLAB (Supplementary Section VII). Using these, we can estimate trap size $\alpha L$ and trap depth $d$ for a given shift $s$. 
 Fig.~\ref{fig:particle-gate-measure}(b) shows the characterization of multiple traps with $n_c = 2, 3, 4$. Here, the trap size $\alpha L$ and normalised depth $d/k_BT$ are plotted as a function of the shift $s$ in blue and red color respectively. These traps are accessible at different glass coverslip locations obtained by moving the horizontal stage of the microscope under the same experimental conditions.
We emphasize that $\alpha L$ and $d$ are effective quantities describing the potential due to surface charges experienced by the particle and not direct measurements of the size or charge density of individual patches on the glass surface. 

\section{Conclusions}
We have demonstrated transport of one-dimensional clusters of colloidal particles along a driven non-equilibrium path. The experiments capture subtle features such as particle position correlations and  shifts in particle positions depending on particle number $n$ as predicted by our theory and simulation results. 
We used well-established wavefront modulation techniques to obtain optical vortices that robustly reproduce the theoretical setting of driven one-dimensional paths experimentally. We were able to characterize deep potential wells impossible to measure at experimental timescales using Kramers' escape rate theory \cite{kastner2011umbrella}. We also propose the utility of these techniques in the study of particle-gates in biological systems that demonstrate a high level of regulation and control. Of particular interest is the ``direct" knock-on mechanism due to Coulomb repulsion between adjacent $\ce{K+}$ ions proposed for channel opening \cite{kopfer2014ion,kratochvil2016instantaneous, mita2021conductance}. Here, larger $\text{K}^+$  ions occupy effective binding sites or traps and start conducting once all the critical binding sites are filled, whereas smaller $\text{Na}^+$ ions see many more traps and don't get selectively conducted. Particle gates in the form of ion channels and nuclear pore complexes
have also been studied in the literature as one-dimensional structures that regulate transport in biological
systems~\cite{shi2018single, bharambe2024cryoem, li2016selective}.
These are highly selective non-equilibrium structures consisting of
rugged potential landscapes with barriers of several hundred $k_BT$,
traversed under a driving force provided by an underlying transmembrane
potential difference~\cite{ghavami2016energetics, gu2020lipid, mita2021conductance}.
To recreate such particle-gates experimentally, nanoscale surface functionalization techniques can be used to engineer charged patches on the surfaces adjacent to the particles as discussed in literature\cite{kalasin2015}. This leads to tuning of the forces and potential wells experienced by the particle and also has applications in particle sorting based on size, along with engineering particle-gates.


\textbf{Acknowledgements:} We would like to thank Tridib Sadhu, Yogeesh Yerrababu and Abhishek Chaudhuri for useful discussions. 
\bibliographystyle{apsrev4-2}

\bibliography{references}
\end{document}


\title{Supplementary Information}
\date{}

\maketitle
\tableofcontents
\pagenumbering{roman}
\subsection*{Supplementary Videos: \href{https://zenodo.org/records/20688804}{Link}}
\begin{itemize}
    \item Supplementary Video 1: A $\SI{4}{\micro\meter}$ diameter colloidal particle moving in a perfect optical vortex without a trap. 
\item Supplementary Video 2: Smaller particles with $\SI{0.6}{\micro\meter}$ diameter getting stuck in many more traps than a larger particle as shown above. 

\item Supplementary Video 3: Multiple particles with $\SI{2}{\micro\meter}$ diameter moving along an optical vortex corresponding to the long exposure photograph shown in main Fig. 2(c) 

\item Supplementary Video 4: Circular trap path unwrapped to a straight line showing particles getting stuck and moving at collective velocities corresponding to data in main Fig. 4(a) 
\end{itemize}
\subsection*{Supplementary dataset for Fig. 4(b): \href{https://zenodo.org/records/20688482}{Link}}
\pagebreak

\section{Results on Tilted cosine Potential}

\begin{figure}[ htbp]
	\centering
	\includegraphics[width=.99\linewidth]{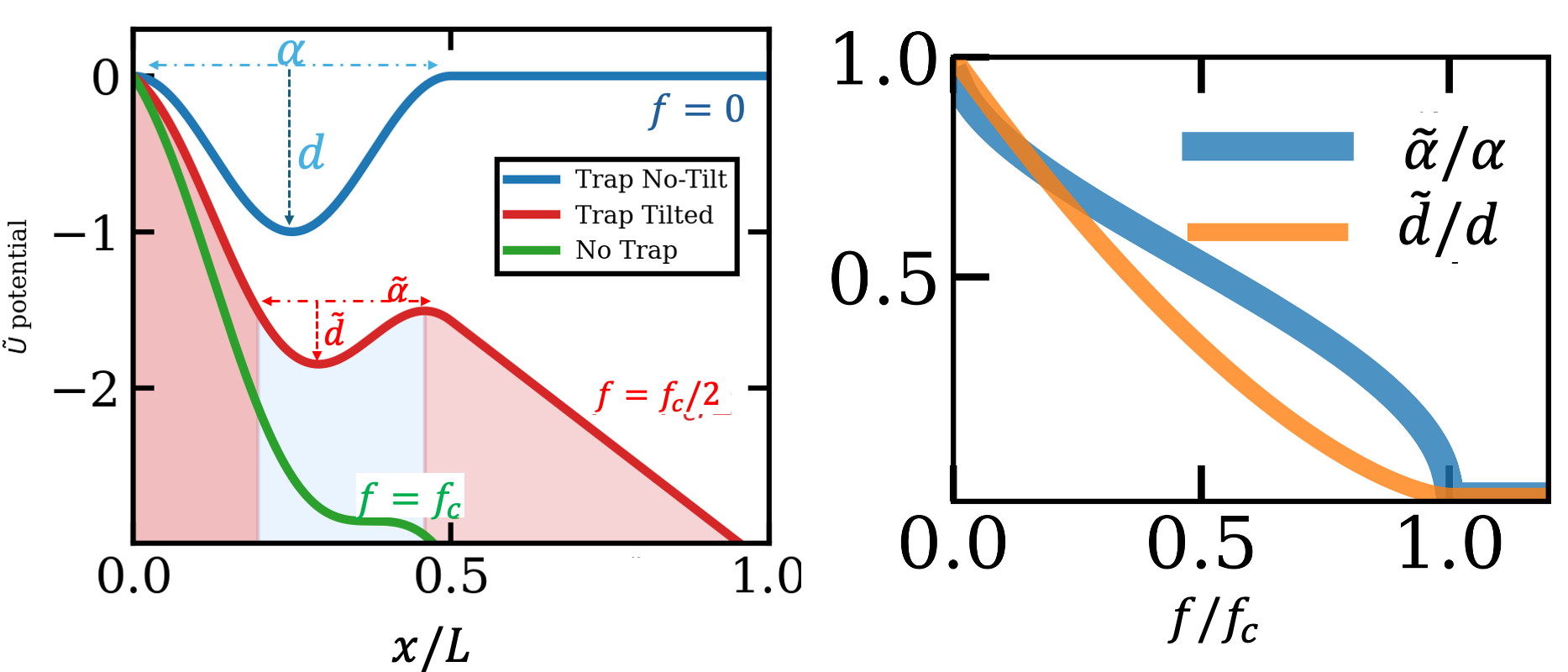}
	\caption{(a) shows the tilted potentials for various values of $f$. (b) shows how the non-equilibrium trap size factor $\tilde \alpha$ and trap depth $\tilde d$ vary with the tilt force.}
	\label{fig:SuppTiltedU}
\end{figure}

Consider a periodic potential well $U$ as defined in the main text, characterized by a size parameter $\alpha$ and depth $d$ along a path of length $L$ as shown in Fig. \ref{fig:SuppTiltedU}(a) represented by $U(x)$ (blue) as
\begin{equation*}
	U(x) = 
	\begin{cases}
		\frac{d}{2}\cos\left(\frac{2\pi x}{\alpha L}\right), & 0 < x < \alpha L \\
		\frac{d}{2}, & \alpha L < x < L
	\end{cases}
\end{equation*}
where $x$ is periodic in  $[0,L] $,   $k_B$ is the Boltzmann constant, $T$ is temperature, and $ \alpha L$ is the size of the trap. In the presence of non-equilibrium driving force $f$ along the path $L$, the `tilted' non-equilibrium potential is given by $ \widetilde{U}(x) = d/2\cos\left(2\pi x/\alpha L\right) - f x$ represented with the red curve in Fig. \ref{fig:SuppTiltedU}(a). As the tilt force $f$ is increased, effective non-equilibrium barrier depth $\widetilde{d}$ and effective size of the trap $\tilde \alpha L $ reduce until $f_c=\pi\,d /\alpha L$ (green curve) at which point the potential well disappears.  Variation of normalized trap depth $\widetilde {d}/d$ and normalised trap size $\widetilde {\alpha} / \alpha$ versus $f_c/f$ is shown in Fig. \ref{fig:SuppTiltedU}(b) inset.

\section{Phase Mask Calculation for Spatial Light Modulators}

\begin{figure}[ htbp]
	\centering
	\includegraphics[width=.99\linewidth]{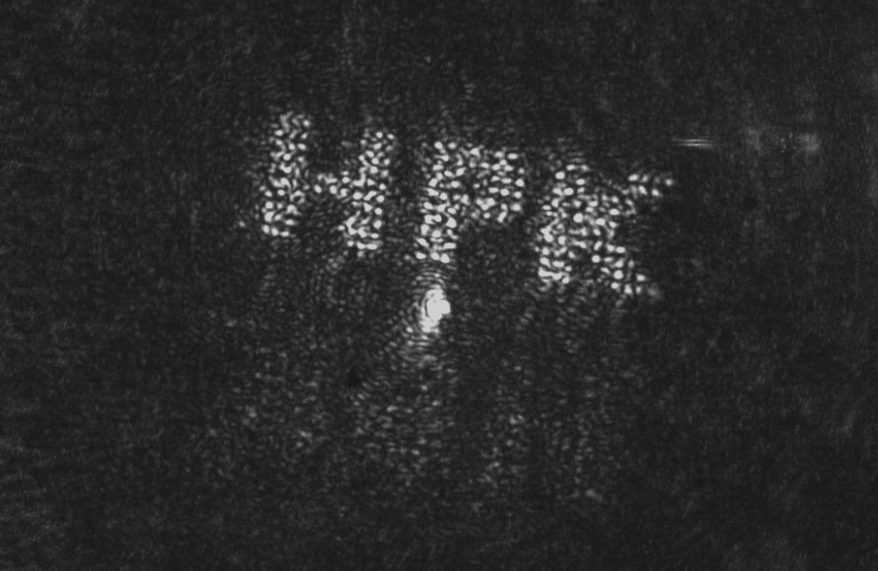}
	\caption{shows an HPK pattern used for calibration.}
	\label{fig:SuppHPK}
\end{figure}

To establish accuracy and calibration of the spatial light modulator, an HPK reference pattern was recreated in the objective plane as shown in Fig. \ref{fig:SuppHPK}. 

A Laguerre--Gaussian (LG) beam of topological charge $\ell$ and radial
index 0 has an electric field in the paraxial approximation of the
form \cite{Allen1992}
\begin{equation}
	U_\ell( \phi, z)
	\propto\,e^{i\ell\phi}\,e^{ikz},
\end{equation}
where $(r, \phi)$ are polar coordinates in the transverse plane,
 and $e^{i\ell\phi}$ is the helical phase factor carrying orbital
angular momentum (OAM) of $\ell\hbar$ per photon.
The topological charge $\ell \in \mathbb{Z}$ sets the OAM content.
The SLM encodes the superposition of the vortex and axicon phases.
In the continuous (analogue) representation, the complex field
transmission of the hologram is
\begin{equation}
	U_\text{SLM}(r,\phi)
	= e^{i\Phi(r,\phi)},
	\qquad
	\Phi(r,\phi) = \ell\phi - \frac{2\pi r}{r_o},
	\label{eq:slm_phase}
\end{equation}
where the first term imprints the helical wavefront and the second term encodes phase
for the axicon. We did trial and error for a few values of $r_o$ to get the desired output of the LG beam ring radius in the focal plane. The combined
phase $\Phi$ is then wrapped to the interval $[0, 2\pi)$ for display
on the SLM:
\begin{equation}
	\Phi_\text{wrapped}(r,\phi)
	= \Phi(r,\phi) \bmod 2\pi
	= \left(\ell\phi - \frac{2\pi r}{r_o}\right) \bmod 2\pi.
	\label{eq:wrapped}
\end{equation}

In pixel coordinates $(i,\,j)$ centred at the pattern centre
$(C_x,\,C_y)$ on the SLM, with pixel pitch $\delta$ (in metres), the
polar coordinates are
\begin{equation}
	r_{ij} = \delta\sqrt{(i-C_x)^2 + (j-C_y)^2},
	\qquad
	\phi_{ij} = \mathrm{arctan2}\!\left(j - C_y,\; i - C_x\right),
	\label{eq:pixelcoords}
\end{equation}
and the phase assigned to pixel $(i,j)$ is
\begin{equation}
	\boxed{
		A_{ij}
		= \left(\ell\,\phi_{ij} - \frac{2\pi\,r_{ij}}{r_o}
		\right) \bmod 2\pi.}
	\label{eq:pixel_phase}
\end{equation}
It is essential to use the four-quadrant inverse tangent
$\mathrm{arctan2}(y, x)$ in equation~(\ref{eq:pixelcoords}) rather
than the standard $\arctan(y/x)$, so that $\phi_{ij}$ spans the full
range $(-\pi, \pi]$ and the helical phase winds continuously around
the full azimuthal circle. Using $\arctan$ alone restricts the output
to $(-\pi/2, \pi/2)$, introducing a phase discontinuity along the
half-plane $i < C_x$ and effectively halving the topological charge.

A liquid-crystal-on-silicon (LCOS) SLM modulates the phase by applying
a voltage-controlled birefringence. The device maps an 8-bit grey level
$g \in [0, 255]$ to a phase $\varphi(g)$ via a calibration look-up
table. For the Hamamatsu LCOS-SLM used here, a full $2\pi$ phase stroke
corresponds to grey levels $0$--$224$ (at $\lambda = 1064$~nm), so
the wrapped phase in equation~(\ref{eq:wrapped}) is converted to grey
levels by
\begin{equation}
	g_{ij}
	= \left\lfloor \Phi_\text{wrapped}(r_{ij},\phi_{ij})
	\times \frac{224}{2\pi} \right\rfloor,
	\label{eq:greylevel}
\end{equation}
where $\lfloor\cdot\rfloor$ denotes the floor function. The resulting
8-bit image was saved as a bitmap and streamed to the SLM. The python/MATLAB programs for the calculation of the SLM pattern can be found at \cite{HarshGithubSLM}.

\section{Geometric overlap and effective intercepting area}
\label{sec:intercept}

\begin{figure}[ htbp]
	\centering
	\includegraphics[width=.99\linewidth]{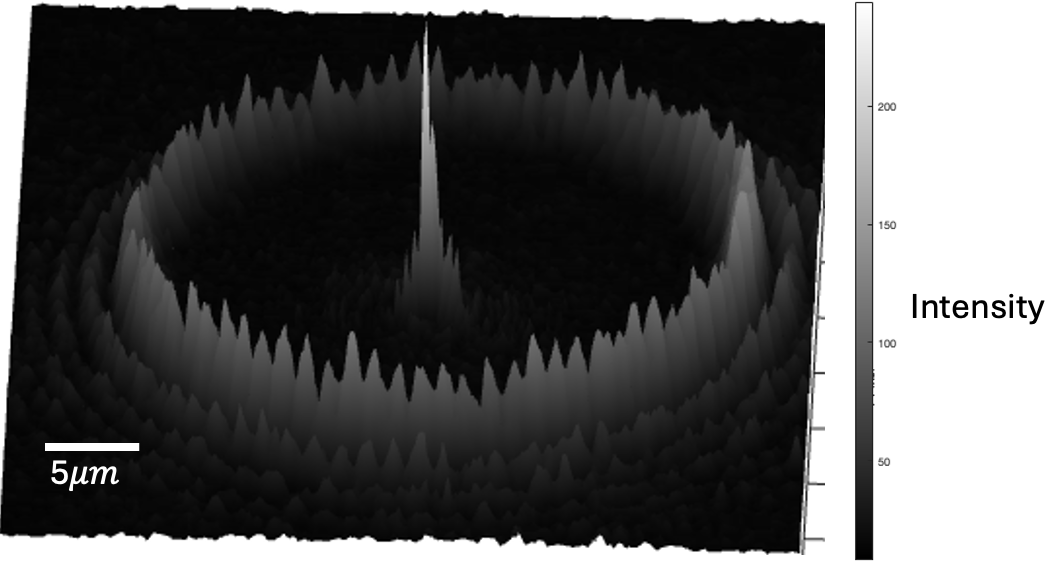}
	\caption{ shows a 3D representation of the image data for the LG beam, used to measure the trap characteristics}
	\label{fig:SuppLGBeam}
\end{figure}
Total intensity of $0.3~\mathrm{W}$ reaches the objective focal 
plane as measured by a power meter. Using the radial profile 
obtained from the image of the LG beam in Fig.~\ref{fig:SuppLGBeam}, 
we estimate that an intensity of light $\sim 0.2~\mathrm{W}$ is 
concentrated along the circular trap, with a trap width 
$w \approx \SI{2}{\micro\meter}$.

A particle of radius $a = \SI{1}{\micro\meter}$ placed at $R\approx \SI{22}{\micro\metre}$ is 
illuminated by the ring intensity over the region of overlap 
between its disk projection and the radial intensity profile. 
Two limiting cases bound the intercepted power:
\begin{enumerate}
	\item If $2a \ll w$, the particle samples a narrow slice 
	of the ring at intensity $I(R)$, and the intercepted area 
	is the geometric cross-section $\pi a^{2}$.
	\item If $2a \gg w$, the particle intercepts the full 
	azimuthal strip of width $w$, and the intercepted area is 
	$A_{\mathrm{strip}} = 2a \cdot w$.
\end{enumerate}
For our parameters, $2a = 2~\mu\mathrm{m}$ is comparable to the 
ring width $w \approx 2~\mu\mathrm{m}$, placing us in the 
intermediate regime where neither limit is exactly correct. 
Numerically integrating the overlap of a disk of radius $a$ 
centred at $R$ against the profile, the effective 
intercepted power is approximately $P_{\mathrm{int}} \approx 
1.6~\mathrm{mW}$ across the cross-section of the particle.

Assuming complete transfer of the optical momentum, this 
intercepted power exerts an axial radiation force
\begin{equation}
	F_{z} = \frac{n_w\,P_{\mathrm{int}}}{c} \sim 10~\mathrm{pN},
\end{equation}
where $n_w \approx 1.33$ is the refractive index of the surrounding 
water. This analysis has also been discussed in \cite{Jain2023Geometry}.

\section{Mie scattering and radiation pressure efficiency}

\label{sec:mie}

The optical momentum intercepted by the particle is not transferred 
with unit efficiency. Most of it is carried forward by elastically 
scattered light and never imparted to the sphere. We quantify this 
with the dimensionless radiation-pressure efficiency $Q_\mathrm{pr}$, 
defined so that the force along the propagation direction is
\begin{equation}
	f_\mathrm{beam} = Q_\mathrm{pr}\,\frac{n_w P_{int}}{c},
	\label{eq:radforce}
\end{equation}
where $P_{int}$ is the intercepted power (Sec.~\ref{sec:intercept}) and 
$n_w = 1.33$ accounts for the momentum of light in water. Following 
Bohren and Huffman \cite{BohrenHuffman1983},
\begin{equation}
	Q_\mathrm{pr} = Q_\mathrm{ext} - g\,Q_\mathrm{sca},
	\label{eq:Qpr}
\end{equation}
where $Q_\mathrm{ext}$ and $Q_\mathrm{sca}$ are the extinction and 
scattering efficiencies and $g = \langle\cos\theta_s\rangle$ is the 
scattering asymmetry parameter. The subtraction in 
Eq.~(\ref{eq:Qpr}) encodes the central effect: forward-scattered 
photons ($g \to 1$) retain their original momentum and exert little 
net pressure, so a strongly forward-scattering particle transfers far 
less momentum than its geometric cross-section would suggest.

For a homogeneous polystyrene sphere we evaluate $Q_\mathrm{ext}$, 
$Q_\mathrm{sca}$, and $g$ from the standard Mie series 
\cite{Mie1908, BohrenHuffman1983}, computed numerically with the 
Wiscombe truncation $n_\mathrm{max} = x + 4x^{1/3} + 2 \approx 18$ 
\cite{Wiscombe1980}. With our parameters
\begin{equation}
	a = 1~\mu\mathrm{m}, \quad \lambda = 1064~\mathrm{nm}, 
	\quad n_p = 1.59, \quad n_w = 1.33,
\end{equation}
the size parameter is $x = 2\pi a n_w/\lambda \approx 7.85$ and the 
relative index is $m = n_p/n_w \approx 1.195$, giving
\begin{equation}
	Q_\mathrm{ext} \approx 2.05, \quad
	Q_\mathrm{sca} \approx 2.05, \quad
	g \approx 0.87, \quad
	Q_\mathrm{pr} \approx 0.27.
\end{equation}
The equality $Q_\mathrm{ext} \approx Q_\mathrm{sca}$ confirms that 
polystyrene is essentially non-absorbing at $1064~\mathrm{nm}$ 
\cite{Sultanova2009}, so all extinction is elastic scattering. The 
strongly forward-peaked scattering ($g \approx 0.87$) suppresses the 
momentum transfer to $Q_\mathrm{pr} \approx 0.27$ -- only about a 
quarter of the geometric-optics expectation. This is similar to the 
$\sim 25\%$ efficiency quoted in the main text.

With $P \approx 1.6~\mathrm{mW}$, Eq.~(\ref{eq:radforce}) gives a force along the beam axis as
\begin{equation}
	f_\mathrm{beam} 
	= 0.27 \times \frac{1.33 \times 1.6\times10^{-3}}{3\times10^{8}}
	\approx 1.9~\mathrm{pN}.
\end{equation}
The helical wavefronts tilt the local momentum azimuthally by 
$\sin\theta = \ell\lambda/(2\pi n_w R) \approx 0.23$ 
($\ell = 40$, $R \approx 22~\mu\mathrm{m}$), so the tangential force is expected to be
\begin{equation}
	f_\mathrm{tan} = f_\mathrm{beam}\sin\theta \approx 0.45~\mathrm{pN},
\end{equation}

\section{Drag force near a planar wall}

For a sphere of radius $a$ translating parallel to a planar 
wall at a surface gap $h$ (the distance between the sphere 
edge and the wall), the bulk Stokes drag $\gamma_0 = 6\pi\eta a$ 
is enhanced by the hydrodynamic interaction with the wall. 
The enhancement factor $\beta = \gamma/\gamma_0$ depends on 
the dimensionless gap $h/a$ and is captured by two 
complementary approximations valid in different regimes. The 
five-term Faxén-type expansion \cite{happel1965low} gives
\begin{equation}
	\gamma_\text{F} = \gamma_0 \left[1 - \frac{9}{16}\frac{a}{b} 
	+ \frac{1}{8}\left(\frac{a}{b}\right)^3 
	- \frac{45}{256}\left(\frac{a}{b}\right)^4 
	- \frac{1}{16}\left(\frac{a}{b}\right)^5\right]^{-1},
	\label{eq:faxen}
\end{equation}
where $b = h + a$ is the centre-to-wall distance and 
$\xi = a/b$. This expansion is formally valid for 
$h \gg a$ and saturates at a finite value 
$\beta_\text{F}(\xi \to 1) = 256/83 \approx 3.08$ as the 
truncated series approaches its formal limit. In the 
physically relevant near-contact regime $h \ll a$, the exact 
asymptotic result of Goldman, Cox and Brenner 
\cite{Goldman1967} for parallel translation in a quiescent 
fluid gives
\begin{equation}
	\gamma_\text{GCB} \approx \gamma_0\left[\frac{8}{15}
	\ln\!\left(\frac{a}{h}\right) + 0.9588\right],
	\label{eq:gcb}
\end{equation}
which diverges logarithmically as $h \to 0$.

Equation~(\ref{eq:gcb}) gives $\beta_\text{GCB}\approx 2.5-3$ 
for a gap $h\approx 30-50 nm$ which indeed matches $\beta_F$. 
The Debye length in deionised water can be expected to be approximately $\lambda_D \approx 100 nm$\cite{israelachvili2011intermolecular}.  For a particle of radius $a = 1~\mu$m held against the coverslip by DLVO electrostatic repulsion, equilibrium 
surface gap of $50 nm$ is reasonably set by the balance 
between the optical axial force and the screened Coulomb 
repulsion force at the salt concentrations relevant to our buffer.

\section{Simulation of particle dynamics and correlated interaction}

To simulate the Langevin equation, we used small timesteps with $dt$ in the range $10^{-3}$ to $10^{-5}$. Smaller $dt$ was used for smaller trap characteristic parameter $\alpha$ which sets the size of the trap. The displacement of the particle in one timestep due to the drift force or potential force scales as $dt$ whereas it scales as $\sqrt{dt}$ due to noise. This made sure that the noise adds as a random walk. Also, the timesteps were much smaller compared to the particle size or the length of the trajectory. The total time steps of the simulation $N$ were kept in the range $10^4$ to $10^6$ corresponding to $dt$ varying from $10^{-3}$ to $10^{-5}$ so that the physical time was set at a constant. 

We modelled steric repulsive interaction between particles using the WCA (Weeks-Chandler-Anderson) potential. 
For a pair of particles
$i$ and $j$ separated by a distance $r = |x_i - x_j|$ along the ring,
the pair potential reads
\begin{equation}
	U_{\mathrm{WCA}}(r) =
	\begin{cases}
		4\varepsilon\!\left[\left(\dfrac{\sigma}{r}\right)^{12}
		- \left(\dfrac{\sigma}{r}\right)^{6}\right] + \varepsilon,
		& r < 2^{1/6}\sigma,\\[6pt]
		0, & r \geq 2^{1/6}\sigma,
	\end{cases}
\end{equation}
where $\sigma = 2a$ is the contact diameter, $a$ is the particle
radius, and $r_c = 2^{1/6}\sigma \approx 1.122\,\sigma$ is the cutoff distance. Here, $\epsilon$ is the well-depth of the WCA potential.
Both $U_{\mathrm{WCA}}$ and its derivative vanish continuously at
$r_c$, so the resulting force is smooth.

The corresponding pair force, directed along the line connecting the
two particles, is
\begin{equation}
	F_{\mathrm{WCA}}(r) = -\frac{dU_{\mathrm{WCA}}}{dr}
	= \frac{24\varepsilon}{r}\!\left[2\!\left(\frac{\sigma}{r}\right)^{12}
	- \left(\frac{\sigma}{r}\right)^{6}\right],
	\qquad r < r_c.
\end{equation}

The contact force at $r=\sigma$ is $12 \epsilon/a$.  We set $\epsilon$ in such a way that the contact force is an order of magnitude higher than the drift force. This, along with a small $dt$ helps prevent particles crossing and maintains single file diffusion. 

Simulations were carried out over High Performance Computing (HPC) cluster and python codes can be found at \cite{HarshGithub_OT}. The simulation time increased rapidly with the number of interacting particles. The simulation required more than a day for the highest number $n=15$. We swept over various values of trap size $\alpha$ varying from 0.005 to 1 as well as the number of particles $n$ varying from 1 to 15. We measured the transport efficiency $\gamma\mu$ in simulations by calculating the average distances travelled by all the particles over simulation time. 
For calculations of the shift of the leading particle or the correlations of positions of particles in the cluster behind the trap, $n$ in the range $2$ to 4 was used while sweeping over the $\alpha$ values.

\section{Shift measurement}

\begin{figure}[ htbp]
	\centering
	\includegraphics[width=.99\linewidth]{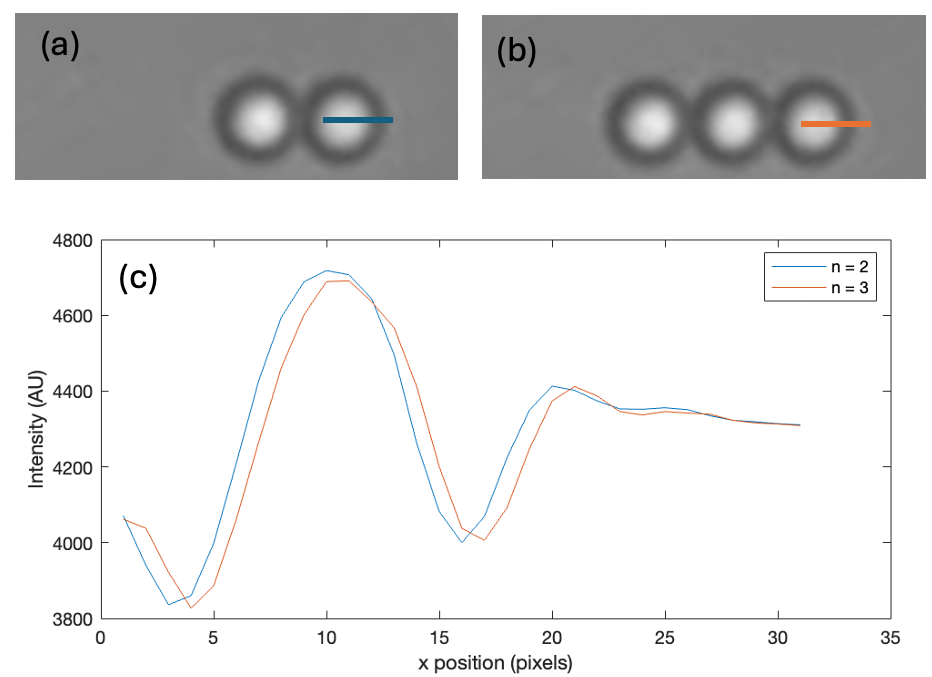}
	\caption{(a-b) show an image of particles in a cluster of size $n=2,3$, (c) shows the shift in intensity along the lines labelled in (a):blue and (b):orange}
	\label{fig:Supp:shift1}
\end{figure}

Fig. \ref{fig:Supp:shift1} shows the slight shift in intensity profile (along the marked blue and red lines in (a-b). The intensity profile shown in (c) shows a shift as the number of particles ($n$) in the cluster is increased from $n=2$ to $n=3$. Fitting this, we can obtain the value for the shift. 

\begin{figure}[htbp]
	\centering
	\begin{subfigure}[t]{0.38\linewidth}
		\centering
		\includegraphics[width=\linewidth]{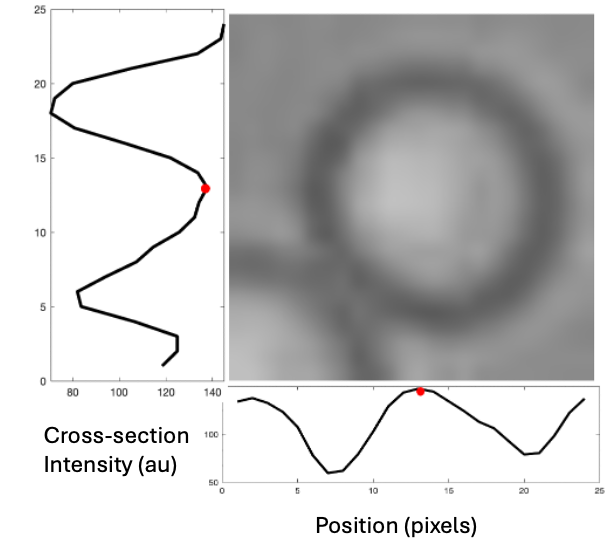}
		\caption{}
		\label{fig:shift23-i}
	\end{subfigure}
	\hfill
	\begin{subfigure}[t]{0.61\linewidth}
		\centering
		\includegraphics[width=\linewidth]{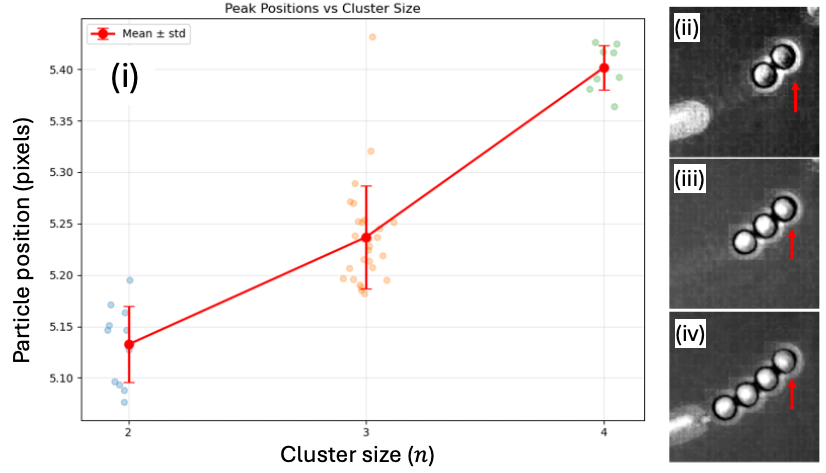}
		\caption{}
		\label{fig:shift23-ii}
	\end{subfigure}
	\caption{(a) shows the intensity profile along the x and y-direction used to determine centre of the colloidal particle. (b) shows the determination of shift in pixels (i) as the position of the leading particle shifts with the cluster size $n$.}
	\label{fig:Supp:shift23}
\end{figure}

Fig. \ref{fig:Supp:shift23} (a) shows the method for determination of the center of the particle with sub-pixel resolution. Correspondingly, the mean position shift of the leading particle (labelled by the arrow in Fig. \ref{fig:Supp:shift23}(ii-iv)) is shown in (b). 
\pagebreak

\bibliographystyle{unsrt}
\bibliography{references}